\documentclass[twocolumn]{aastex62}

\received{}
\revised{}
\accepted{}

\usepackage{amssymb}
\usepackage{amsmath}
\begin{document}
\title{A Cancellation Nanoflare Model for Solar Chromospheric and Coronal Heating}
\correspondingauthor{P. Syntelis}
\email{ps84@st-andrews.ac.uk}
\author{E.R. Priest}
\affiliation{St Andrews University\\
Mathematics Institute\\
St Andrews KY16 9SS\\
UK}
\author{L.P. Chitta}
\affiliation{Max Planck Institute for Solar System Research\\
377077, G\"{o}ttingen\\
Germany}

\author{P. Syntelis}
\affiliation{St Andrews University\\
Mathematics Institute\\
St Andrews KY16 9SS\\
UK}


\begin{abstract} 
Nanoflare models for heating the solar corona usually assume magnetic 
braiding and reconnection as the source of the energy. However,  recent observations  at record spatial resolution from the {\sc Sunrise} balloon mission suggest that photospheric magnetic flux cancellation is much more common than previously realised. We therefore examine the possibility of three-dimensional reconnection driven by flux cancellation as a cause of chromospheric and coronal heating. In particular, we estimate how the heights and amount of energy release 
produced by flux cancellation depend on flux size, flux cancellation speed  and overlying field strength.  
\end{abstract}


\section{Introduction} 
\label{sect_1}
 Many interesting proposals have been 
put forward to solve the major puzzle of how the solar atmosphere is heated,  including
 MHD waves and magnetic reconnection  \citep[e.g.,][]{klimchuk06,parnell12,priest14a}. However, the mechanisms have not yet been conclusively identified. The classic picture for  {\it nanoflares} invokes magnetic braiding of footpoints to create many  current sheets 
that dissipate by reconnection throughout the corona \citep{parker88}. This was later developed into the coronal tectonics model \citep{priest02a}, with dissipation at separatrix surfaces. 

\begin{figure*}
\begin{center}
\includegraphics[width=\textwidth]{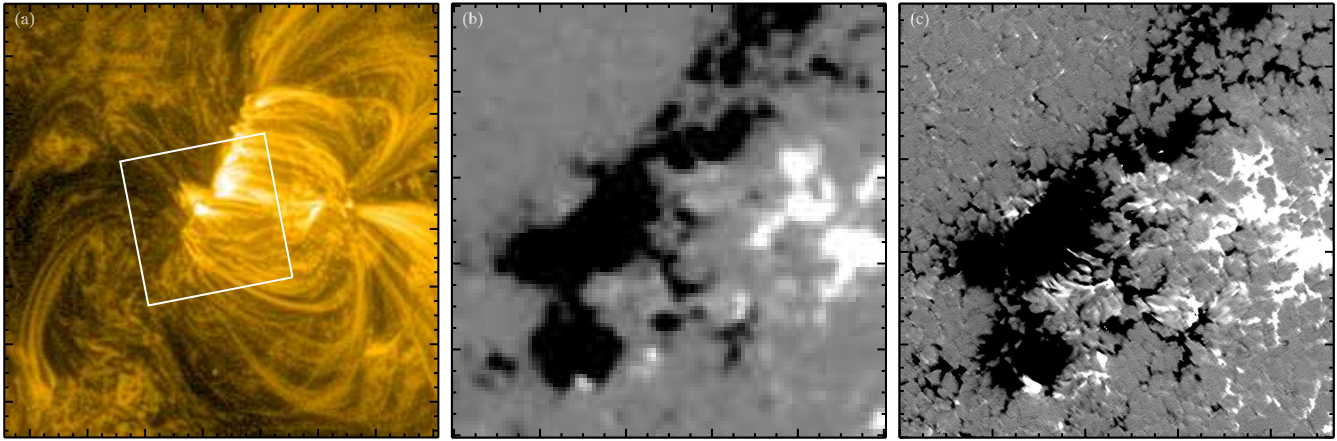}
\caption{A coronal image of an active region on 2013 June 12 at 23:45\,UT, and the underlying magnetic field. (a) An image from the SDO/AIA 171\,\AA\ filter in a  $150\arcsec\times150\arcsec$ field of view. (To improve the contrast of the image, we have used a multi-scale Gaussian normalisation technique \citep{morgan14}). The white box covers an area of $51\arcsec\times51\arcsec$ and encloses footpoint regions of several coronal loops. (b) SDO/HMI magnetogram showing the distribution of the photospheric line of sight magnetic field for the white box region of panel (a). (c) Same as (b) but for the {\sc Sunrise}/IMaX observations. The magnetic flux density is saturated at $\pm250$\,G. See \citet{chitta17a} for further details.}
\label{fig1}
\end{center}
\end{figure*}
Observing the evolution of magnetic field patches at high spatial resolution near the footpoints of magnetic loops may be crucial to understanding chromospheric and coronal heating. 
Recently, the IMaX instrument (Imaging Magnetograph eXperiment) on two flights of the  {\sc Sunrise} balloon Mission 
\citep{solanki10a,solanki17b} has revealed glimpses 
of the photospheric magnetic field at much 
higher spatial resolution than before, namely, 0.15 arcsec, a factor of six better 
than the Helioseismic and Magnetic Imager (HMI) instrument on the Solar Dynamics Observatory (SDO) \citep{pesnell12}. 
Using the observations from the first flight of {\sc Sunrise}, \citet{smitha17} tracked magnetic features with fluxes of $10^{15}$--$10^{18}$ Mx in the Quiet Sun and found a flux emergence and cancellation
rate of 1100 Mx cm$^{-2}$ day$^{-1}$. This rate is an order of magnitude higher than previous measurements. \citet{chitta17a} observed the footpoints of extreme ultraviolet (EUV) loops (171 \AA) in a new active region. At SDO/HMI resolution (1 arcsec) they appeared to be simple bipolar regions, with the loops joining two unipolar regions. However, higher-resolution maps at 0.15 arcsec (100 km) from IMaX on {\sc Sunrise} revealed
mixed magnetic polarity at the loop footpoints, with flux cancellation at a rate of $10^{15}$ Mx s$^{-1}$  (Fig. \ref{fig1}).

It is well known that flux cancellation can liberate magnetic energy through reconnection. The  general relevance of such flux events and associated reconnection for chromospheric and coronal energetics certainly needs further scrutiny. Indeed, three other pieces of  evidence support the possible importance of 
flux cancellation for chromospheric and coronal heating. Firstly, the 
heating of coronal loops may often be focussed near their feet 
\citep[e.g.,][]{priest00a,aschwanden08}.  Secondly, the well established  view that at least X-ray bright points are produced 
mainly by flux cancellation is supported  by observations 
 \citep[e.g.,][]{martin85,falconer99} and theory  
 \citep{priest94b,parnell95,longcope98,parnell04,archontis14b}. Thirdly, the driving of magnetic reconnection  by flux emergence or cancellation has different observational consequences depending on the location in height of the reconnection, which in turn depends on the  magnitudes of the  flux source and the overlying field strength (see Section \ref{sect_2}). Thus, energy release  can produce: low down in the atmosphere around sunspots or in the Quiet Sun an Ellerman bomb  in the wings of H$\alpha$ \citep{rouppe16,hansteen17}; in the chromosphere of an active region UV bursts (or IRIS bombs) \citep{peter14b}; in the transition region explosive events \citep{brueckner83,innes11}, blinkers \citep{harrison97}; and in the corona  X-ray bright points and X-ray jets \citep{shibata92,shimojo07}.
 
Recent studies further emphasize the possible, wide-spread role of reconnection during flux cancellation as the source of coronal loop brightenings. \cite{tiwari14} and \citet{huang18} discussed examples of flux cancellation triggering coronal brightening in apparently braided  loops. \citet{chitta17b} observed that coronal loops in an evolved active region respond to an underlying ultraviolet burst and bidirectional jets, which in turn are triggered by magnetic reconnection at heights of 500 km above the photosphere driven by magnetic interactions leading to flux cancellation. Furthermore, \citet{chitta18}  observed flux cancellation near the footpoints of coronal loops hosting nanoflares in the core of an active region. They identified complex mixed polarity field at the loop footpoints, where flux was cancelling at a rate of 10$^{15}$ Mx s$^{-1}$. Plasma at 1 MK in 171 \AA\ showed fluctuations at one footpoint where flux cancellation was occurring and a steady evolution at the other footpoint. By comparing the energy content of the loop with that of the magnetic energy below the chromosphere (where reconnection is presumed to take place), they concluded that the analysed flux cancellation events provide sufficient energy to heat the corona to temperatures exceeding 5 MK.

The realisation that there is very much more photospheric 
flux cancellation than previously thought leads us to consider flux 
cancellation as a possible cause of chromospheric and coronal heating.
We  present
some theoretical aspects (Section \ref{sect_2}) and conclude with a discussion 
(Section \ref{sect_3}), in which the height and amount of energy release are estimated as functions of the flux and overlying field strength.

\section{Theory for Energy Release at a Reconnecting Current Sheet} 
\label{sect_2}
\begin{figure*}[!htbp]
    {\centering \includegraphics[width=12cm]{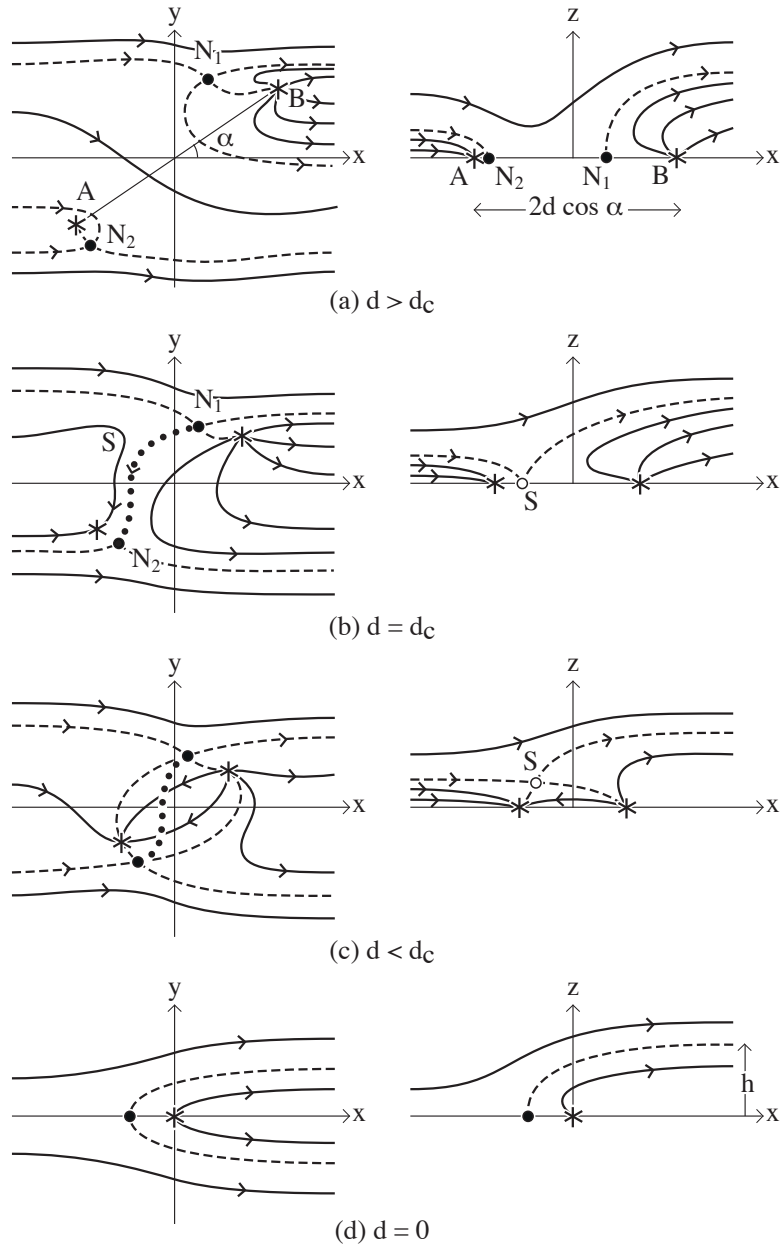}
    \caption{The magnetic field structure seen from above 
    (left) and from the side (right) during the approach along the line AB of 
    oppositely directed photospheric  sources  (stars) of flux 
    $-F_p$ and $F (\geqslant -F_p)$, separated by a distance $2d$ and situated in an 
    overlying uniform horizontal magnetic field ($B_{0}$). Separatrix magnetic field lines are dashed, other magnetic field lines are solid curves, null points are large
    solid dots and a separator (S) is a curve of dots seen from above and
    an unfilled dot when seen from the side. (a) is for $d>d_{c}$,  (b) is for 
    $d=d_{c}$, when a separator  first appears lying in $z=0$, (c) is for 
   $d<d_{c}$ when the separator arches above the surface, and (d) is for  $d=0$.}
\label{fig2}}
\end{figure*}
Here we make some theoretical estimates of the energy release by 
steady-state magnetic reconnection in three dimensions, using  basic theory  \citep{priest14a} and developing it in 
new ways. We  calculate the 
rate of magnetic energy conversion when flux cancellation drives 
reconnection as two oppositely directed photospheric magnetic sources approach and 
cancel in an overlying field that is for simplicity here assumed horizontal \citep{stenflo13,Orozco12}. Inclined fields will be treated in future.

 \subsection{Basic Properties of the  Configuration}
\label{sect_{2.1}}
Suppose a photospheric source of negative parasite polarity of flux $-F_p$ 
lies next to a larger source ($ F\geqslant -F_p$) of  positive polarity. The sources are at  points A $(-d \cos \alpha,-d \sin \alpha)$ and B $(d \cos \alpha,d \sin 
\alpha)$,  a distance $2d$ apart in the $xy$-plane and inclined at an angle $\alpha$ 
to the direction of an overlying field of strength $B_0$.
Consider what happens as they approach one another at speeds $\pm v_{0}$ along the line joining A to B.  

The  magnetic field above the 
photosphere ($y>0$) is 
\begin{equation}
{\bf B}=\frac{F\ \bf{\hat{r}_{1}}}{2\pi r_{1}^{2}}-\frac{F_p\ \bf{\hat{r}_{2}}}{2\pi r_{2}^{2}}+B_0{\bf{\hat{x}}},
\label{eqn6}
\end{equation}
where
	\begin{eqnarray}
	   {\bf r}_{1}=(x-d\cos \alpha){\bf \hat{x}}+(y-d \sin 
\alpha){\bf \hat{y}}+z{\bf \hat{z}},  \\
	   {\bf r_{2}}=(x+d\cos \alpha){\bf \hat{x}}+(y+d \sin 
\alpha){\bf \hat{y}}+z{\bf \hat{z}},
	 \end{eqnarray}
are the vector distances from the two sources to a point P($x,y,z$).

Fig.\ref{fig2} sketches the evolution of the topology of the magnetic field in the horizontal
$xy$-plane and the vertical $xz$-plane. 
Consider what happens when the distance ($2d$) between the two sources 
decreases from a large value. When the sources are so far apart that $d>d_{c}$, say,
then there are two separatrix surfaces (containing two null points 
N$_{1}$ and N$_{2}$) that completely surround the 
fluxes that enter A and leave B, so that no flux links 
A to B (Fig.\ref{fig2}a).  On the other hand, when $d=d_{c}$, a 
separator bifurcation occurs in which these two 
separatrices touch at a separator field line (S) that lies in the photospheric plane ($z=0$) and joins the two null points (Fig.\ref{fig2}b).  
Furthermore, when $d<d_{c}$, the separator 
rises above $z=0$  and a new domain is 
created bounding magnetic flux that passes under S
and links source A to source B (Fig.\ref{fig2}c).  Finally, when
$d=0$, the  parasitic flux has completely cancelled, leaving a separatrix surface that encloses the flux from the remaining dominant polarity (Fig.\ref{fig2}d).

\begin{figure*}
    {\centering \includegraphics[width=12cm]{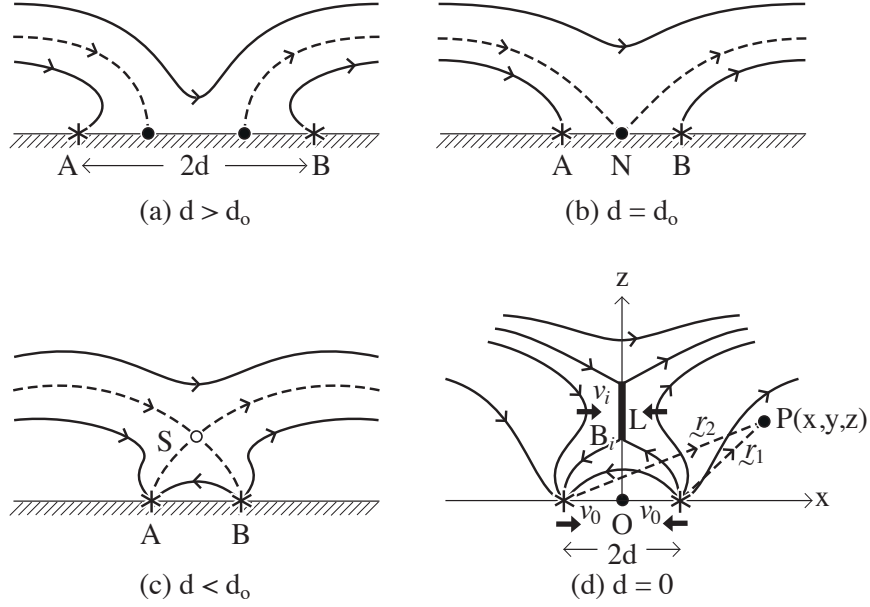}
    \caption{Magnetic topologies in the vertical $xz$-plane when 
    $\alpha = 0$ and $-F_p=F$ for (a) $d>d_{0}$,  (b) 
    $d=d_0$ and (c) $d<d_0$ in terms of the interaction distance 
    ($d_0=[F/(\pi B_0)]^{1/2}$). 
    (d) shows the notation when there is a 
    reconnection region of length $L$ at which the inflow velocity 
    and magnetic field are $v_i$ and $B_i$.}
\label{fig3}}
\end{figure*}

Here we focus on the particular case of equal flux sources ($-F_p=F$) with $\alpha =0$ (Fig.\ref{fig3}), so that the line joining the flux sources is aligned 
with the overlying magnetic field and we can take the analysis 
much further while retaining the main physics.

\subsection{Equal Flux Sources Aligned with Overlying Field ($-F_p=F$, $\alpha=0$)}
\label{sect_{2.2}}

Consider in detail what happens as the flux sources approach one another as 
$d$ decreases. A natural length-scale for the configuration  is the {\it interaction distance} \citep{longcope98}
\begin{equation}
d_{0}=\left(\frac{F}{\pi B_{0}}\right)^{1/2}.  \nonumber
\end{equation}

When $d>d_0$, there is no flux
connecting the sources (Fig.\ref{fig3}a)
and two first-order null points lie  on the $x$-axis between the 
sources.  When $d=d_0$, there is a local bifurcation in which the nulls combine to give a high-order null at the origin (Fig.\ref{fig3}b). When $d<d_{0}$, a new semicircular separator is born in the $yz$-plane and its intersection with the 
$xz$-plane (marked S in Fig.\ref{fig3}c)
rises along the $z$-axis to height $z_{S}$, say, so that 
magnetic flux now lies under the separator and connects the sources. 
The magnetic field is axisymmetric about the $x$-axis 
and so the separator is actually a semi-circular ring of null points at distance $z_{S}$ from the origin in every plane through the $x$-axis. 
In the case of unequal flux sources, the separator becomes a field line joining two nulls in the $xy$-plane, as shown in Fig.\ref{fig2}c.

Along the $z$-axis, ${B}_{y}={B}_{z}=0$ and 
\begin{equation}
 \frac{{B}_{x}}{B_0} = 
 -\frac{{d}\ d_0^{2}}{({d}^{2}+{z}^{2})^{3/2}}+1.
 \label{eqn10}
\end{equation}
The location (${z}={z}_{S}$) of the null where the field 
vanishes is therefore given  by 
\begin{equation}
{z}_{S}^{2}={d}^{2/3}{d_0}^{4/3}-{d}^{2}    
\label{eqn11}
\end{equation}
and is sketched in Fig.\ref{fig4}a as a function of $d$. When
${d}=d_0$, the null is located at the origin, and, as 
${d}$ decreases it rises along the $z$-axis to a maximum of 
$(z_S)_{max}=(4/27)^{1/4}d_0$ at ${d}=(1/3)^{3/4}d_0$.  Thereafter, the null 
falls back to the origin as ${d}\rightarrow 0$.

The maximum height of the null point varies with $B_0$ and $F$, as shown in Fig.\ref{fig4}b. 
The height is typically about  $0.6 d_0$, and so it lies in the chromosphere when $F$ is small enough or $B_0$ large enough. 
As the flux sources approach, the null point rises from the photosphere to its maximum height and then falls, but the energy that is released may spread to larger heights along the separatrix field lines that link to the reconnection site.

Note that in the more general case where the magnitudes of the two fluxes are not equal, when all the parasitic polarity flux has cancelled we are left with the situation shown in Fig.\ref{fig2}d. Here the flux from the remaining major polarity reaches a maximum  height $ h$, say, to which the field line from the null point asymptotes. It may be estimated from the equation of the field line in the plane $y=0$ through the null $(-1/\sqrt 2,0,0)$, namely,
\begin{equation}
{\textstyle{\frac{1}{2}} z^2} - \frac{{\textstyle{\frac{1}{2}} x}\ d_0^2}{( z^2+ x^2)^{1/2}}=\textstyle{\frac{1}{2}}d_0^2.
\nonumber 
\end{equation}
Thus, as $ x\rightarrow \infty$ on that field line, $ z \rightarrow  h$ and we find  $h= \sqrt 2 d_0$.

\subsection{The Input Plasma Speed ($v_{i}$) and Magnetic Field 
($B_{i}$) at the Reconnection Region}
\label{sect_{2.3}}
When analysing flux cancellation, the natural parameters, for each value of the 
source separation ($2d$), are the interaction distance ($d_{0}$), the flux source speed ($v_{0} 
\equiv {\dot d}\equiv dd/dt$) and the overlying field strength ($B_{0}$).
We now therefore proceed to calculate the inflow 
speed ($v_{i}$) and magnetic field ($B_{i}$) to the current sheet and 
the sheet length ($L$) as functions of those parameters for fast reconnection.

First of all, consider $B_{i}$. If the potential field near a null point has the form $B_{x}=kz$,
then, when a current sheet forms, the magnetic field at the inflow to
the sheet becomes  $B_{i}={\textstyle{\frac{1}{2}}}kL$.
Thus, after using Eqn.(\ref{eqn10}) to find $k$, we obtain
\begin{equation}
\frac{B_{i}}{B_{0}}=\frac{3[1-(d/d_{0})^{4/3}]^{1/2}}{2(d/d_{0})^{1/3}}\frac{L}{d_{0}}.   
\label{eqn13}
\end{equation}

\begin{figure}[h]
{\centering 
     \includegraphics[width=7cm]{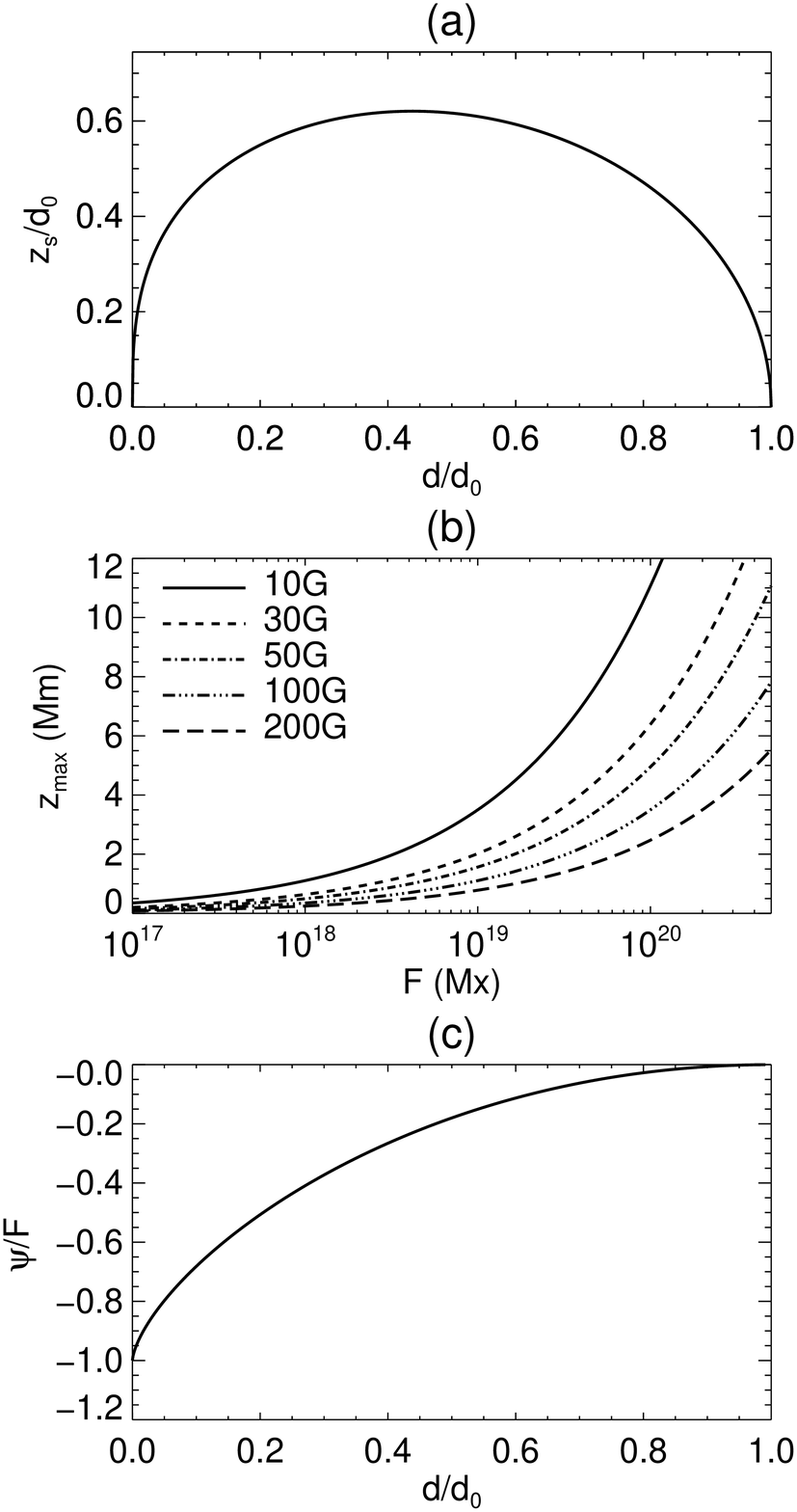}
\caption{(a) The height ($z_{S}$) of the null point in 
Fig.\ref{fig3}c as a function of the half-separation ($d$) 
    between the two sources, where $d_{0}=[F/(\pi B_{0})]^{1/2}$ is the interaction distance.
(b) The maximum height of the null point as a function of the flux $F$  for different values of the overlying magnetic field $B_0$.
(c) The flux ($\psi$) below the null point as a function of $d$.}
\label{fig4}}
\end{figure}

Next, consider $v_{i}$, which may be calculated from the rate of 
change (${\dot \psi} \equiv d\psi /dt$) of magnetic flux through the 
surface  bounded by the $y$-axis and a
semicircle of radius $z_{S}$ out of the plane of Fig.\ref{fig3}c.
After using Faraday's Law and $\bf E+\bf v \times \bf B=\bf 0$, this rate of change of flux becomes
 \begin{equation}
   \dot{\psi}\equiv \frac{d\psi}{dt} = -\pi z_{S}E  =  \pi z_{S}v_{i}B_{i}.
\label{eqn14}
  \end{equation}
However, $\psi$ may be calculated from the magnetic flux below 
$z_{S}$ through the surface, namely,
\begin{equation}
\psi= \int_{0}^{z_{S}}\pi z\ B_{x}\ dz=F\left[\frac{3}{2}\left(\frac{d}{d_0}\right)^{2/3}-\frac{1}{2}\left(\frac{d}{d_0}\right)^{2}-1\right].
    \nonumber
\end{equation}
This is sketched in Fig.\ref{fig4}c, from which it can be seen that, as expected, the reconnected flux 
vanishes when $d=d_{0}$ and increases monotonically in magnitude to $F$ 
as the separation ($2d$) between the sources approaches zero. 

After differentiating $\psi$ with respect to $t$, we find
\begin{equation}
\dot \psi= \frac{v_{0}F}{d_0} \left[\left(\frac{d}{d_0}\right)^{-1/3}- \frac{d}{d_0}\right].
\label{eqn15}
 \end{equation}   
Then, after substituting into Eqn.(\ref{eqn14}) for $\dot \psi$ from this equation, for
$z_{S}$ from Eqn.(\ref{eqn11}) and for $B_{i}$ from Eqn.(\ref{eqn13}),  the required expression for $v_i$ becomes
\begin{equation}
 {v_{i}}=\frac{2v_{0}}{3}\frac{d_{0}}{L}\left(\frac{d_0}{d}\right)^{1/3}.
  \label{eqn16}
\end{equation}

\subsection{Energy Release by Fast Reconnection}
\label{sect_2.4}
Three possibilities have been studied for fast reconnection, all of which may occur within our model, depending on the microscopic plasma physics at work. 
Firstly, according to Petschek or Almost-Uniform reconnection theory \citep[e.g.,][]{priest14a}, the internal structure of the reconnection region consists of a central small sheet and four slow-mode shock waves, at which most of the energy conversion takes place, with $0.4$ of the inflowing magnetic energy being converted to heat. Secondly,  collisionless reconnection is aided by the Hall effect, when the resistive diffusion region is replaced by an ion diffusion region and a smaller electron diffusion region, but the same fast maximum rate of reconnection  results \citep[e.g.,][]{shay98a,birn07}. Thirdly, when the central sheet is long enough, it goes unstable to secondary tearing mode instability and a regime of impulsive bursty reconnection results \citep[e.g.,][]{priest86b,loureiro07}, with a mean energy conversion and reconnection rate similar to the other cases.

The rate of inflow of magnetic energy from one side at speed $v_i$ and with field $B_i$ and density $\rho_i$ through a surface with height $L$ and extending a distance $\pi z_S$ along the current sheet at the separator is just the Poynting influx
$(E H_i L\pi z_S=E B_i L\pi z_S/\mu)$. However, the magnitude of the electric field is  $E=v_iB_i$, and an equal amount of magnetic energy flows in from the other side of the sheet, so  the total rate of conversion of energy to heat  from both sides is
\begin{equation}
    \frac{dW}{dt}=0.8\frac{v_{i}B_{i}^{2}}{\mu}L\pi z_{S}.
\label{eqn17}
\end{equation}

$L$ is  determined  by  the condition that the inflow speed  $v_i=\alpha v_{Ai}$, where $\alpha$ is typically between 0.01 and 0.1, and  $v_{Ai}=B_{i}/\sqrt{\mu \rho_i}$ is the inflow Alfv\'en speed.
Then, after setting $\alpha v_{Ai}=\alpha (B_{i}/B_{0})v_{A0}$ in Eqn.(\ref{eqn16}), where  $v_{A0}=B_{0}/\sqrt{\mu \rho_i}$ is a hybrid Alfv\'en speed, and using Eqn.(\ref{eqn13}), we obtain
\begin{equation}
    \frac{L^{2}}{d_{0}^{2}}=\frac{4v_{0}}{9\alpha 
    v_{A0}}\frac{1}{[{1-(d/d_{0})^{4/3}}]^{1/2}}.
\label{eqn19}
\end{equation}
After substituting for $B_{i}/B_{0}$ from Eqn.(\ref{eqn13}), $v_{i}/v_{A0}$ from Eqn.(\ref{eqn16}), and $L$ from Eqn.(\ref{eqn19}), the rate of 
energy conversion Eqn.(\ref{eqn17}) becomes 
\begin{equation}
    \frac{dW}{dt}=0.8\frac{2\pi}{3}\frac{v_{0} B_{0}^{2}}{\mu}d_{0}^{2}\frac{M_{A0}}{\alpha}\frac{[1-(d/d_0)^{4/3}]}{(d/d_0)^{2/3}}
\label{eqn20}
\end{equation}
for a given flux source speed $v_{0}$, overlying field $B_0$, interaction distance $d_0$, 
Alfv\'en Mach number ($M_{A0}=v_{0}/v_{A0}$) and source separation $2d$.

\section{Discussion} 
\label{sect_3}
Inspired by the remarkable {\sc Sunrise} observations, we here propose that magnetic reconnection driven by photospheric flux cancellation may be a ubiquitous mechanism for powering coronal loops and also for releasing heat in the chromosphere.  We suggest the outlines of a theoretical model for the interaction between two opposite-polarity sources of flux $\pm F$ in an overlying horizontal field $B_0$, which can be greatly developed in future by  sophisticated computational experiments.

Three key roles are played by the {\it interaction distance} which may be written
\begin{equation}
d_{0}=6  \left(\frac{F_{19}}{B_{1}}\right)^{1/2}\ {\rm Mm},  \nonumber
\end{equation}
where $F_{19}$ is the flux in units of $10^{19}$ Mx and $B_{1}$ is the overlying field in units of 10 G. The first is that, as the opposite polarity sources approach one another, they drive reconnection as soon as $d<d_0$. 
For example, small flux sources of $10^{17}$ Mx  give values for $d_0$ of 0.6 Mm  in a 10 G field or 0.2 Mm  in a 100 G field. On the other hand, large flux sources of  $10^{20}$ Mx, give values of  19 Mm in a 10 G field  and 6 Mm in a 100 G field.

The second role is to determine the maximum height ($(z_S)_{max}\approx 0.6\ d_0$) for the reconnection location and so explain why flux cancellation sometimes leads to energy release in the photosphere, sometimes in the chromosphere and sometimes in the transition region or corona. Thus, $(z_S)_{max}$ lies in the photosphere if $F<2\times 10^{17}$ Mx for  $B_0=10$ G or $F<2\times 10^{18}$ Mx for  $B_0=100$ G. On the other hand it lies in the chromosphere if $2\times 10^{17} < F < 3\times 10^{18}$ Mx for  $B_0=10$ G or $2\times 10^{18} < F < 3\times10^{19}$ Mx for  $B_0=100$ G. These computed maximum reconnection heights are consistent with those from magnetic field extrapolations for  chromospheric bursts \citep[e.g.,][]{chitta17b,tian18}. 

The third role for $d_0$ is that, when the overlying field is horizontal, the height reached by the field lines that link to the reconnection site varies between  $1.4\ d_0$ when reconnection starts and $1. 5\ d_0$ when the reconnection height peaks (at $(z_S)_{max})$). Thus, we expect energy to propagate down towards the photosphere and up to a height of $1.4 - 1.5 \ d_0$.  This lies purely within the photosphere and chromosphere when $F<4\times 10^{17}$ Mx for  $B_0=10$ G or $F<4\times 10^{18}$ Mx for  $B_0=100$ G.  Of course the height will be much larger when the field lines are inclined to the solar surface.

Next, consider the energy liberated. In order to heat the Quiet-Sun chromosphere and corona, we need $4\times 10^6$ and $3\times 10^5$ erg cm$^{-2}$ sec$^{-1}$, respectively, whereas in an active region the corresponding needs are $2\times 10^7$ and $10^7$ erg cm$^{-2}$ sec$^{-1}$, respectively.  Let us evaluate the rate of heat produced in the chromosphere from Eqn.(\ref{eqn20}) with typical values of $d=0.4\ d_0$ and $\alpha=0.1$ \citep{priest14a}. Then the expression (\ref{eqn20}) may be written
\begin{equation}
\frac{dW}{dt}=5\times 10^{22}\ v_4\ B_1\ F_{18}\ M_{A0}\ {\rm erg\ /sec},
\nonumber
\end{equation}
where $v_4$ is $v_0$ in units of $10^4$ cm/sec. Thus, for example, in the Quiet Sun, if we adopt values of $v_0=$1 km/sec, $F=10^{18}$ Mx, $B_0=10$ G, $M_{A0}=0.1$ \citep{priest14a}, so that an area of $l_0^2$ is swept out in a time of, say, $10^3$ sec, where $l_0=10^8$ cm, then the heating per unit area is

\begin{equation}
\frac{1}{l_0^2}\frac{dW}{dt}=5\times 10^{6}\ {\rm erg\ cm^{-2}\ s^{-1}},
\nonumber
\end{equation}
which is sufficient to heat the Quiet-Sun chromosphere.  On the other hand, a flux of $F=10^{19}$ Mx and an overlying field of $B_0=100$ G with $M_{A0}=0.01$, characteristic of active regions would give a corresponding value of $5\times 10^{7} {\rm erg\ cm^{-2}\ sec^{-1}}$, which is sufficient for the active-region chromosphere. In turn, if 10--20\% of this leaks through to higher levels, it would be sufficient to heat the corona.   

We have proposed a ubiquitous way of creating nanoflares near the base of chromospheric and coronal loops with sufficient energy to power the chromosphere and corona, building on previous flux cancellation theory \citep[e.g.,][]{parnell95,welsch06}. In future, it will be interesting to develop the model further by means of computational experiments, in order to investigate the nature of the energy release and its propagation along  magnetic loops from the reconnection source.

\acknowledgements
The authors are most grateful for invaluable discussions with Hardi Peter and Clare Parnell.  
L.P.C. received funding from the European Union's Horizon 2020 research and innovation programme under the Marie Sk\l{}odowska-Curie grant agreement No.\,707837. 
The German contribution to {\sc Sunrise} and its reflight was funded by the Max Planck Foundation, the Strategic Innovations Fund of the President of the Max Planck Society (MPG), DLR, and private donations by supporting members of the Max Planck Society.
The Spanish contribution was funded by the Ministerio de Econom\'{i}a y Competitividad under Projects ESP2013-47349-C6 and ESP2014-56169-C6, partially using European FEDER funds. 
The HAO contribution was partly funded through NASA grant number NNX13AE95G. SDO data are courtesy of NASA/SDO and the AIA, and HMI science teams.


\end{document}